\documentclass[aps,pre,showpacs,twocolumn,floatfix]{revtex4-1}

\usepackage{graphicx,color}
\usepackage{amsmath,amssymb}

\newcommand{\potcntact}{%
\begin{figure}[htbp]
   \centering     
   \includegraphics[width=.5\textwidth]{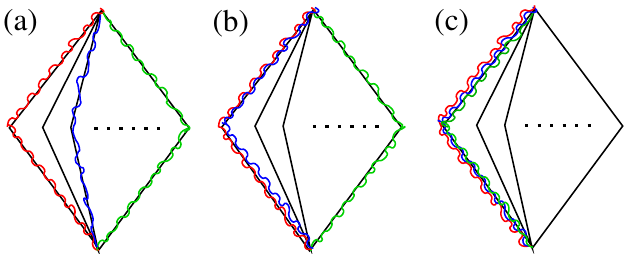}
   \caption{ Different configurations of a triple-stranded DNA are
     shown by the wavy lines on a $d$ dimensional lattice for general $b$.
     (a) Polymers are separate on the path and the total number of
     configurations is $b(b-1)(b-2)$.  (b) Two polymers among three
     are together and the total number is $b(b-1)$.  (c) Three
     polymers are together and the total number is $b$.}
   \label{fig:ptcnt}
 \end{figure}
}
\newcommand{\sglattice}{%
\begin{figure}[htbp]
   \centering     
   \includegraphics[width=0.30\textwidth]{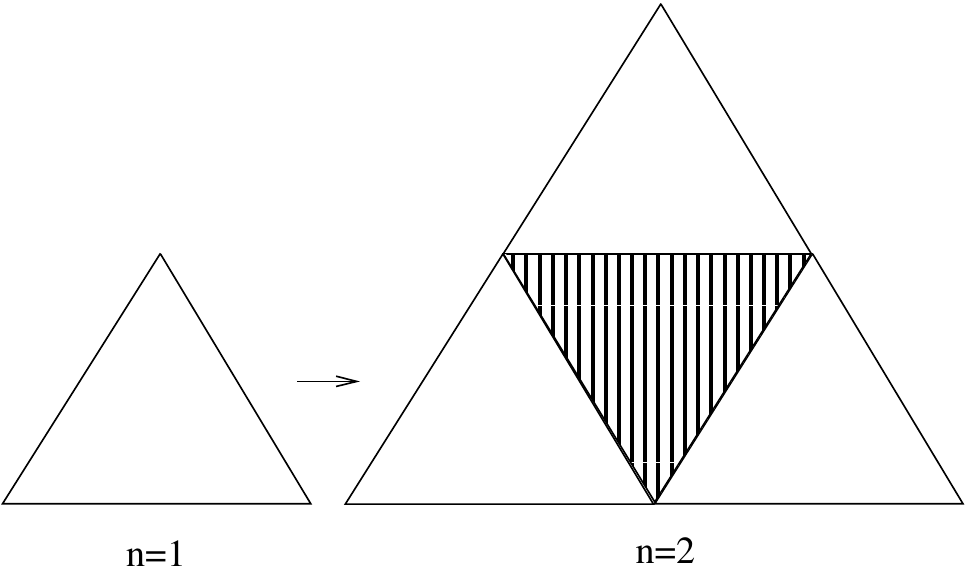}
   \caption{ Schematic diagram of a Sierpinski Gasket for generation
     $n=0,\ 1$.  The basic motif is shown in (a).  In next step shown
     in (b) three triangles are glued together with a forbidden region
     at the center.  }
   \label{fig:latsg}
 \end{figure}
}
\newcommand{\lattsch}{%
\begin{figure}[htbp]
   \centering
   \includegraphics[width=0.35\textwidth]{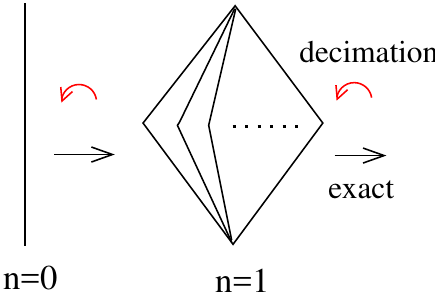}
   \caption{Schematic diagram of an hierarchical lattice.  At each 
    stage a bond is replaced by a basic motif (shown $n=1$).    
    The left arrow denotes the direction of decimation of RG. }
   \label{fig:lat-sch}
 \end{figure}
}
\newcommand{\lattschsg}{%
\begin{figure}[htbp]
   \centering
   \includegraphics[width=0.50\textwidth]{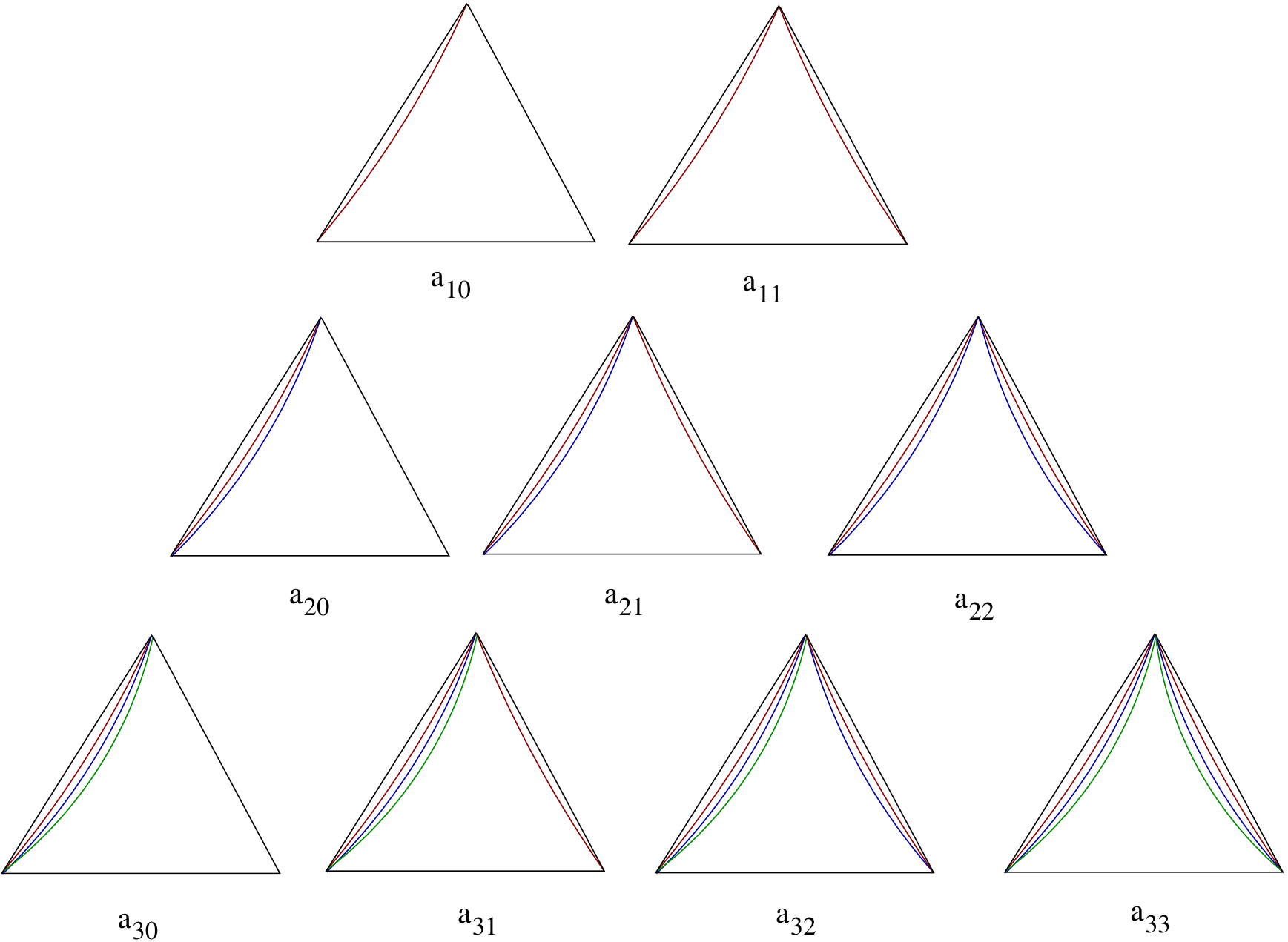}
   \caption{ Possible configurations for single, double and triple
     chain systems.  All possible configurations are represented by
     the partition functions $a_{ij}$ ($i, j=0,1,2,3$).  }
   \label{fig:lat-schsg}
 \end{figure}
}
\newcommand{\mixedsch}{%
\begin{figure}[htbp]
   \centering
   \includegraphics[width=0.45\textwidth]{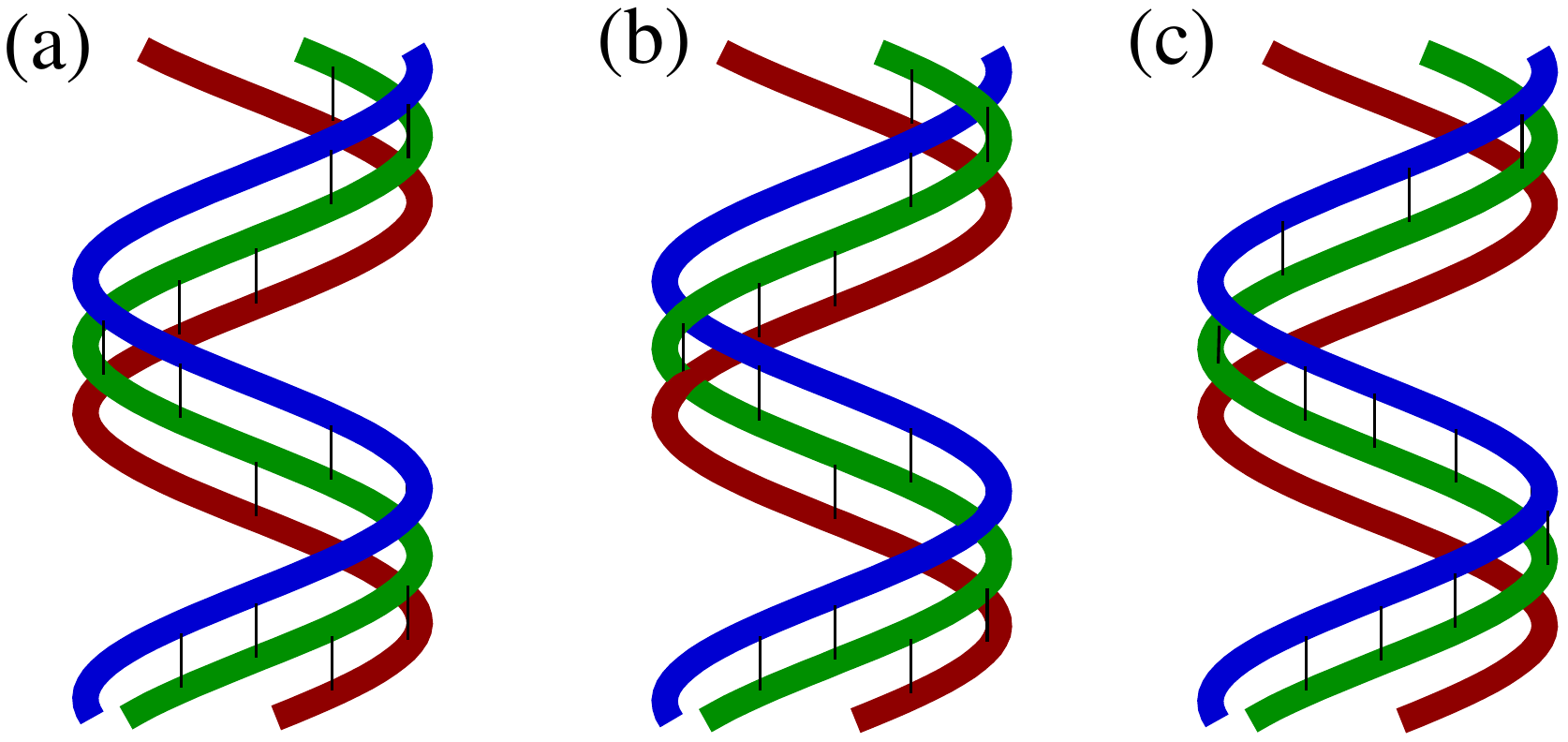}
   \caption{Schematic diagram of the mixed phase of three polymers.
     Per unit length of the chain two monomers are in contact leaving
     the third free.  The pair interaction is shown by the vertical
     bonds.  (a) Polymers cannot cross each other.  (b) Polymers can
     cross each other. (c) Polymers cannot cross each other. Two bound
     and one free strands.  }
   \label{fig:mix-sch}
 \end{figure}
}

\newcommand{\discuss}{%
\begin{figure}[htbp]
   \centering
  \includegraphics[width=0.4\textwidth]{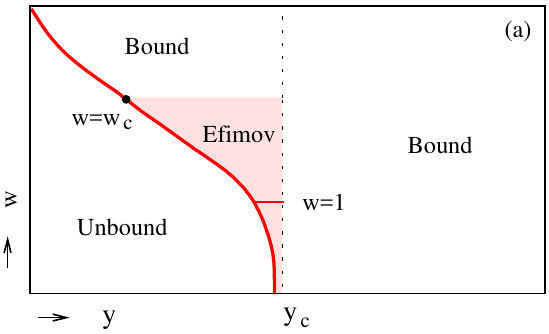}
\hfill\includegraphics[width=0.4\textwidth]{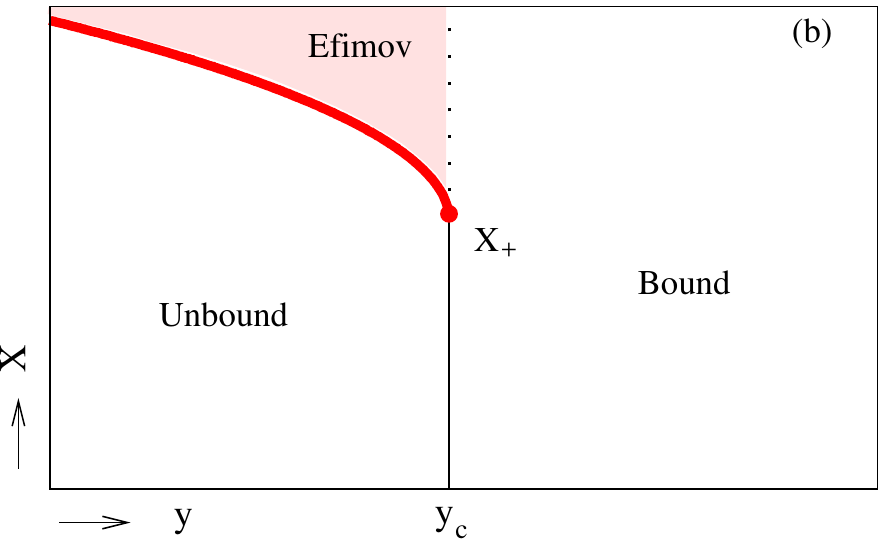}
\hfill\includegraphics[width=0.4\textwidth]{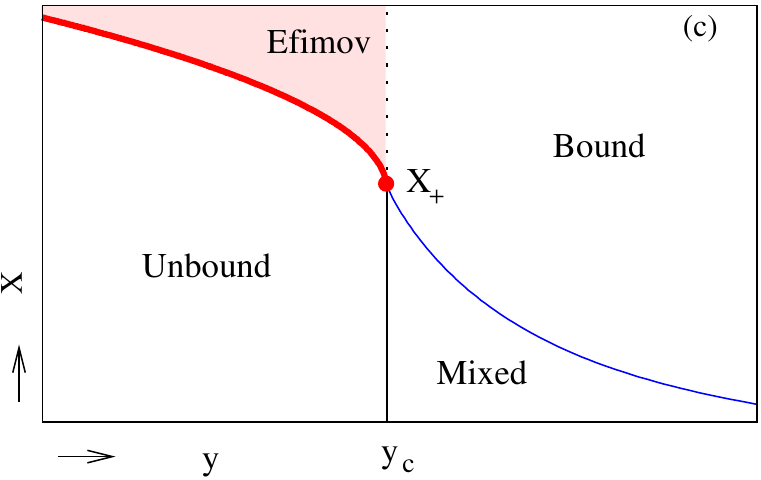}

\caption{Schematic phase diagrams in the $w$-$y$ or the $X$-$y$ plane.
  (a) For $b=8 (d=4)$.  There are only two phases three-strand bound
  state and unbound DNA, separated by the solid line. This is a first
  order transition line.  The $y=y_c$ dotted vertical line is the
  two chain melting line and does not exist for the three chain
  system.  The filled circle on the solid line at $y=1, w=w_c$ is the
  pure three chain melting point.  The horizontal line $w=1$
  corresponds to the traditional Efimov case with pure two body
  interaction.  The Efimov region is the region enclosed by (i) the
  $w=w_c$ line, (ii) the solid line, and (iii) the vertical $y=y_c$
  line.  (b) The $X$-$y$ phase diagram for $b=9\, (d=4.1699)$.  The
  Bound phase now melts via the Efimov state for larger $X$, but at
  $y=y_c$ for smaller $X$.  This phase diagram is similar to that in
  Ref.~\cite{owcz}.  (c) The $X$-$y$ phase diagram for $b=16 (d=5)$.
  There is now a mixed phase on the $y>y_c$ side.  The thin lines
  denote continuous transitions, while the melting from the Efimov
  side (thick solid line) is first order.  The dotted line, a relic of
  the two chain melting, does not exist in the three chain system.
  The two continuous lines and the first order melting line meet at
  the multicritical point at $X_{+}$ at $y=y_c$.}
   \label{fig:dis}
 \end{figure}
}

\newcommand{\mixedsgnc}{%
\begin{figure}[htbp]
   \centering
   \includegraphics[width=0.45\textwidth]{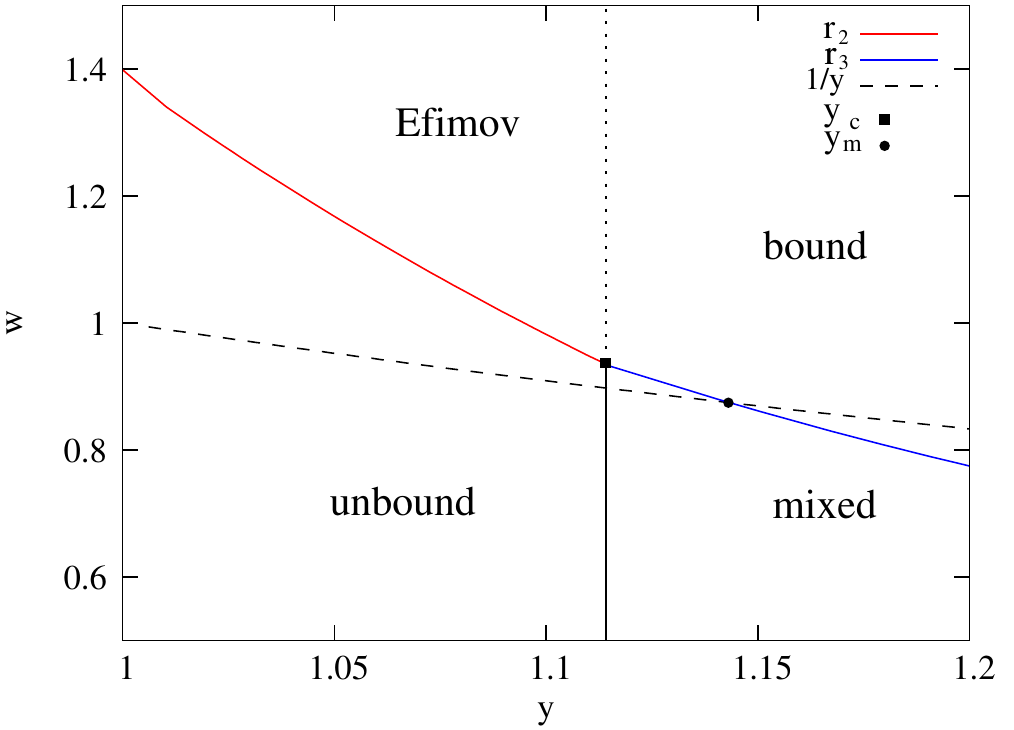}
   \caption{ Phase diagram in the $w$-$y$ plane for the non-crossing
     case with $\sigma=1$.  The Efimov (red) and the mixed (blue)
     phase boundaries intersect the vertical line $y=y_c$.  The dashed
     line is for the TS1 model. }
   \label{fig:mix-sgnc}
 \end{figure}
}

\newcommand{\mixedsgc}{%
\begin{figure}[htbp]
   \centering
   \includegraphics[width=0.45\textwidth]{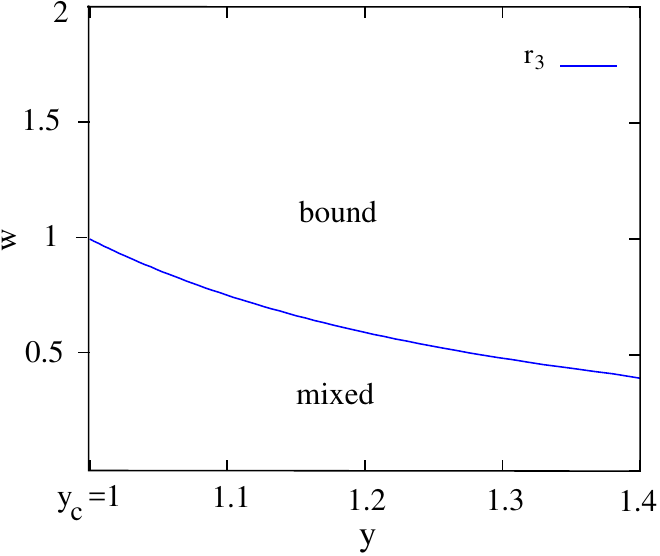}
   \caption{ Phase diagram in the $w$-$y$ plane for the crossing case
     with $\sigma=1$.  Two chain melting is at $y=y_c$.  The phase
     boundary is for the transition from the bound to the mixed phase.
     No two chain or pure three chain melting at finite temperature as
     $y_c=1$ ($w_c=y^{-1}_c=1$). }
   \label{fig:mix-sgc}
 \end{figure}
}

\newcommand{\mixedsgp}{%
\begin{figure}[htbp]
   \centering
   \includegraphics[width=0.45\textwidth]{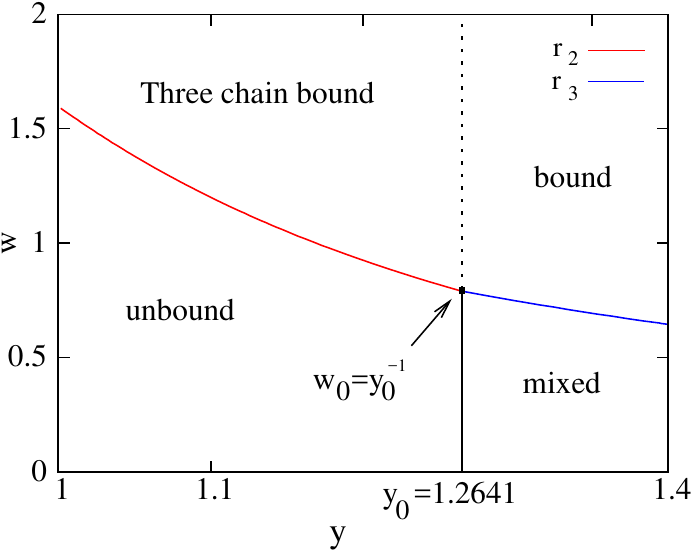}
   \caption{ Phase diagram in the $w$-$y$ plane for the crossing case
     with $\sigma=0$.  All the phases occur in this case like in the
     non-crossing case.  Three phases (three-chain bound, all bound
     and the mixed phase by continuation) coexist at the triple point
     ($y_0=1.264..$, $w_0=y^{-1}_0$).  }
   \label{fig:mix-sg1}
 \end{figure}
}

\newcommand{\sigyw}{%
\begin{figure}[htbp]
   \centering
 \includegraphics[width=0.50\textwidth]{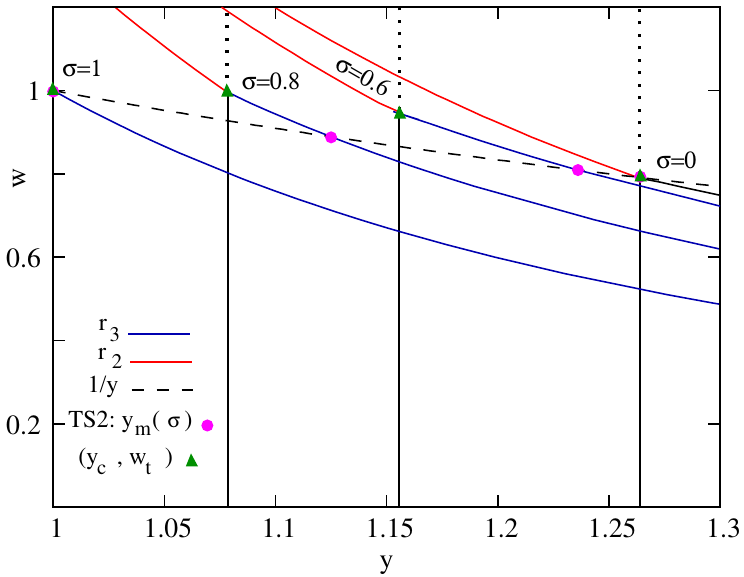}
 \caption{Crossing case.  Phase diagram in the $w-y$ plane for
   different $\sigma$($=1, 0.8, 0.6, 0$).  For each $\sigma$, the thin
   red line is the phase boundary for the unbound to the Efimov region
   of the triplex bound state, where as the blue phase boundary (thick
   line) is for the bound to the mixed state.  These two boundaries
   meet at the filled triangles on the vertical lines at the two chain
   melting $y=y_c(\sigma)$, which is also the melting line of the
   mixed phase.  Filled triangles are the triple points.  The dashed
   curve $w=1/y$ intersects the mixed phase boundaries at
   $y_m(\sigma)$ (shown by filled circles), and reproduces the phase
   diagram of TS2 model\cite{mixed} in the $\sigma$-$y$ plane.  The
   triangle and the circle coincide only at $\sigma=0$ and $\sigma=1$.
  }
   \label{fig:3dplot}
 \end{figure}
}
\begin{document}

\title{Bubble-bound state of triple-stranded DNA: Efimov Physics in
  DNA with repulsion }

\author{Jaya Maji} 
\email{jayam@iiserb.ac.in} 
\affiliation{Department of Physics,  Indian Institute of Science  Education and Research, Bhopal 462 066, India}
\author{Flavio Seno} 
\email{flavio.seno@pd.infn.it}
\author{Antonio Trovato}
\email{antonio.trovato@pd.infn.it}
\affiliation{INFN, Dipartimento di Fisica e Astronomia `Galileo Galilei', Universit\`a di
             Padova, Via Marzolo 8, 35131 Padova, Italy}
\author{Somendra M. Bhattacharjee}
\email{somen@iopb.res.in}
\affiliation{ Institute of Physics, Bhubaneswar 751005, India}

\begin{abstract}
  The presence of a thermodynamic phase of a three-stranded DNA,
  namely, a mixed phase of bubbles of two bound strands and a single
  one, is established for large dimensions ($d\geq 5$) by using exact
  real space renormalization group (RG) transformations and exact
  computations of specific heat for finite length chains.  Similar
  exact computations for the fractal Sierpinski gasket of dimension
  $d<2$ establish the stability of the phase in presence of repulsive
  three chain interaction.  In contrast to the Efimov DNA, where three
  strands are bound though no two are bound, the mixed phase appears
  on the bound side of the two chain melting temperature.  Both the
  Efimov-DNA and the mixed phase are formed due to  strand exchange
  mechanism.
\end{abstract}
\date{\today}
\maketitle

\section{Introduction}
DNA serves as a primary unit of heredity and contains all information
necessary for the living systems.  This complex object is made of two
complementary strands via Watson-Crick base pairing.  One important
milestone of modern biology was the discovery of DNA double-helical
structure\cite{sinden}.  The subsequent discovery of the triple helix
has attracted great attention in medical sciences in view of its
possible use in mapping chromosomes\cite{moser}, inhibition of gene
expressions\cite{grigoriev,gctu}, gene therapy and gene
targeting\cite{knauert}, inducing site-specific mutations(targeted
mutagenesis)\cite{gwang}, interfering with DNA replication\cite{mguo}
and other applications\cite{ajain,mduca,conde}.  Triple helix is
formed by a duplex binding with a single strand DNA, or RNA, or PNA,
via Hoogsteen or reverse Hoogsteen hydrogen bonding for
DNA\cite{rich,yli}, or even by three RNA's\cite{holland,frankkam}.

Recent theoretical studies identified two different types of 
triple-stranded DNA (tsDNA) states near the duplex melting point.  We may
recall that melting of DNA is the phenomenon of temperature induced
separation of double-stranded DNA into single strands.  One of the two
states of tsDNA is a loosely bound state of three strands when no two
are bound\cite{efidna,zeros,mixed,tan1,tan2,mura}, and the other one
is a phase of bubbles of pairwise bound and a single strand below the
melting point.  Both the states are stabilized by strand exchange of a
single strand, and, in a sense, produced by the bubble fluctuations
near the duplex melting point.  The former one, occurring at or above
the melting point on the unbound side, resembles the well known Efimov
effect \cite{efimov1,efimov2} in quantum mechanics, and is called the
Efimov-DNA.  It is not a pure phase but a continuation of the three
chain bound state.  In contrast, the other phase, which we may call a
bubble-bound or a mixed phase, occurs on the bound side, below the
melting point.  It was found as a genuine phase in low dimensional
models with specially tuned interactions, and it differs from the
Efimov-DNA as the chains are locally pairwise bound, but without any
direct three chain contact.

Our aim in this paper is to explore the possibility of the
bubble-bound phase, also called a mixed phase, in triple-stranded DNA
in a wider context.  We would call this a bubble-bound state as a
single strand is always accompanied by a bubble of duplex DNA.  This
paper establishes the existence of the mixed phase for large
dimensions, with $d=5$ as an example.  As already mentioned, this
phase was predicted from fractal lattice studies with $d<2$
\cite{mixed}.  We also study the role of three chain repulsion in
these low dimensional system.

The Efimov effect originates from quantum fluctuations at zero energy
bound states of two attracting nonrelativistic particles.  In a three
particle system with pairwise critical potential, an effective long
range $1/r^2$ attraction appears between two particles at distance $r$
due to the wide excursions of the third particle in the classically
forbidden regions\cite{efimov1,efimov2,braaten}.  Consequently, there
are an infinite number of bound states.  The original continuous scale
invariance in the two particle problem changes to a discrete scale
invariance characterizing the bound states.  Recent experiments have
more or less established the quantum Efimov effect in ultracold atoms,
though the existence of the infinite number of states at the threshold
is yet to be established\cite{expt1,expt2,expt3}.  The triplex bound
state of DNA at the melting point of duplex DNA is the thermal
analogue of the Efimov physics where the denatured bubbles play the
role of classically forbidden paths in quantum mechanics.  A polymer
scaling analysis, using hyperscaling at the dsDNA melting point,
recovers the inverse-square interaction very naturally.  Such
arguments show the importance of the fluctuations of bubbles near the
melting point and the polymer correlations along the single strands of
the bubbles\cite{efidna}.  The Efimov physics has since been extended
to various other systems, like Efimov-driven transition in many body
systems\cite{krauth}, in quantum magnets\cite{magnet}, and in one
dimensional systems under long range interactions\cite{oned}.  Several
studies have probed the importance of dimer-atom states near Efimov
resonance\cite{atomdimer}, and possible topological origins of the
Efimov physics\cite{ueda}.  In this scenario, triple-stranded DNA
appears as a classical testing ground for Efimov physics, which can
also be enriched by other thermal effects and relevant interactions.
In fact, the polymer scaling analysis\cite{efidna} is one of the
simplest ways to see the emergence of the long range interaction at
the heart of the Efimov physics.  The bubble-bound phase we study in
this paper is an analogue of the atom-dimer phase in the quantum
problem.  In the polymer context, such states appear as a stable
thermodynamic phase.

To study the DNA problem in various dimensions, we adopt the model of
three-stranded DNA on diamond hierarchical lattices and Sierpinski
gaskets.  These lattices are constructed in an algorithmic way by
iterative replacement of bonds by a basic motif at each level.  The
discrete scaling symmetry makes these lattices affordable, in contrast
to the regular lattices, for exact calculations or for Renormalization
Group studies of different models
\cite{sgflavio,critpheno,knez,elez,itzy,smsmb1,smb3,berker,lyra1,ac}.
Beside the understanding of complex physical system, hierarchical
lattices also exhibit a lot of interesting mathematics.  The method of
construction of the chosen lattices allows one to express the
partition functions as recursion relations which either can further be
iterated for physical properties in the large lattice limit, or can be
used to decimate for renormalization group (RG) flows and fixed points
(fp).  RG is considered a valuable tool to understand the emergent
behavior of systems with diverging length scales.  By integrating out
the small length scale fluctuations and rescaling, the same system is
regenerated but with renormalized parameters.  The consequent RG
transformations lead to fixed points and separatrices.  The fixed
points represent states of the system where it shows scale invariance
under rescaling of lengths.  A given system need not be at the fixed
points but from the flow patterns of the parameters, one identifies
the transitions and the phases of the system. Moreover, the
transitions can be corroborated by finite size scaling analysis of
finite length thermodynamic properties, like specific heat or energy,
which can be computed exactly in a recursive manner.

In low dimensions, DNA melting does not occur unless a bubble weight
factor is introduced\cite{mixed,mura}.  With such weight factors,
called cooperativity factors, assigned at junctions of bound and
unbound states (Y-fork), the Efimov state and the mixed phase were
studied in different classes of DNA models on the Sierpinski Gasket of
dimension $d_f=\ln 3/\ln 2$.  A detailed study of the Efimov state,
using the fixed size transfer matrix approach, was done in
Ref.~\cite{mura} for 1+1 dimensional Euclidean lattice without any
three chain interaction.  Here also the cooperativity factor was
needed for a melting transition, but the occurrence of the Efimov
state turned out to be sensitive to how the weights are assigned in a
three chain system. A model of three noncrossing chains in $1+1$
dimensions, without any cooperativity factor but with a three chain
interaction, was solved exactly in Ref.~\cite{owcz}.  The phase
diagrams obtained in Ref.~\cite{owcz} resemble partly the phase
diagrams reported here, but without any mixed phase.  In this paper we
focus on the mixed phase which occurs on the bound side of the two
chain melting.  As a continuation of the higher dimensional studies to
$d<2$, we explored the possibilities of the existence of the bubble
bound mixed phase in presence of a three chain repulsion for DNA
models defined on the Sierpinski Gasket.

The paper is organized as follows.  In Sec.~\ref{model1} we consider a
few simplified polymer models to study the Efimov state and mixed
phase of DNA in a lattice of dimension $d>2$ using the renormalization
group approach.  The existence of a phase transition from the
bubble-bound phase to the triple bound state is also shown by exact
specific heat computations.  In Sec.~\ref{model2} we discuss the model
on a Sierpinski gasket of dimension $d<2$ by the method of exact
calculation. Here we extend the analysis of Ref.~\cite{mixed} by
including a three body interaction. In Sec.~\ref{modeldis} we conclude
that the mixed phase is well established from both the lattice models.

\section{Model: $d>2$}\label{model1}
The native base pairing interaction of a DNA is best expressed as
directed polymers on a lattice.  Each monomer of the strands
represents a collection of bases interacting with the monomer of same
sequence index of the other strand as per the Poland-Scherega scheme.
 
The melting of thee-stranded DNA has been studied by real space
renormalization group which can be implemented exactly on hierarchical
lattices of dimensions $d>2$ \cite{efidna,zeros,smsmb1,smb3,salinas}.
The procedure to construct the lattice is to start with a single bond,
at generation $n=0$, and then replace the bond by a diamond motif at 
$n=1$ (See Fig.~\ref{fig:lat-sch}).

The dimension of the lattice is defined by $d=\ln{\lambda
  b}/{\ln\lambda}$, where $\lambda$ is the length scale factor and $b$
is the branching factor connecting the bottom and the top of the
lattice as in Fig.~\ref{fig:lat-sch}.  Parameter $b$ can be tuned to
change the dimension of the lattice.  In our models, $\lambda=2$.

\lattsch

Let us consider three directed polymers being laid from the bottom to
the top of the lattice with no restriction on intersections and
crossings.  See Fig.~\ref{fig:ptcnt}.  Two weight factors are needed
for the DNA problem, viz., $y (=e^{\beta\epsilon})$ associated with
two polymers sharing the same bond with energy $-\epsilon$ and $w
(=e^{-\beta\eta})$, associated with three polymers sharing the same
bond with an additional triplet energy $\eta$.  The triplet
interaction between monomers is attractive if $\eta<0$, i.e., $w>1$ or
repulsive if $\eta>0$, i.e. $w<1$.  Here $\beta=1/T$ is the inverse
temperature (with the Boltzmann constant $k_B=1$), and all the
pairwise interactions are taken to be the same.

Since temperature $T$ is absorbed in $y$ and $w$, for  given 
interaction energies among the  chains we get a temperature curve
in the $y$-$w$ plane, parameterized by $T$.
As a result the $T$ dependence 
of a given set of chains, i.e., for fixed $\epsilon,\eta$,  
can be obtained from the intersection of such a  curve with 
the phase transition lines in the $y$-$w$ plane.
In view of this,  we keep $y$ and $w$ to be 
independent variables.  A combination variable $X=wy^2$ is also found
to be useful, as explained below.


\potcntact
 
\subsection{Renormalization Group equations}
\label{sec:renorm-group-equat}

The RG recursion relations for the weights of the DNA are given
by\cite{efidna,zeros}  
\begin{subequations}
\begin{eqnarray}
y'&=&\frac{b-1+y^2}{b},\label{eqy}\\
w'&=&\frac{(b-1)(b-2)+3 (b-1)y^2+w^2y^6}{b^2 y'^3}.\label{eqw}
\end{eqnarray}
\end{subequations}
These are the RG relations with $w', y'$ denoting the renormalized
values.  The important point is that the renormalization of the two
chain interaction is not affected by the three chain interaction.

The melting point of a
duplex DNA is described by the unstable fixed point (fp) of Eq.~(\ref{eqy})
at $y_c=b-1$ while the high temperature unbound phase is given by the
stable fixed point at $y=1 (T=\infty)$,
These two chain fp's  lead to  different
possibilities for $w$ as 
\begin{subequations}
\begin{equation}\label{pointt}
w^*=\left\{
\begin{array}{lll}
1,  b^2-1, \infty &{\rm for}& y=1\\
w_{-},w_{+}, \infty &{\rm for}& y=y_c\equiv b-1,
\end{array}
\right.
\end{equation}
with
\begin{equation}
  \label{eq:1}
  w_{\pm}=\frac{b^2\pm\sqrt{4-24b+32b^2-12b^3+b^4}}{2(b-1)^3},
\end{equation}
\end{subequations}
being real for $b>b_c=8.56...$. The dimension corresponding to this
special $b_c$ is $d_c=\ln(2b_c)/\ln 2$.  Of these, the fp at
$w_c=b^2-1$ for $y=1$ is the three chain melting point due to a pure
three body interaction. These fixed points for $w$ were discussed
previously in connection with the Efimov-DNA.  We focus here on a new
set of fp's that are found for low temperatures.

\subsection{Zero temperature fixed points: A dilemma}
\label{sec:zero-temp-fixd}
In addition to the above-mentioned fp's, there is a stable fp at
$y=\infty$ that describes the bound state of the duplex at zero
temperature.  In this limit, the RG equation, Eq.~(\ref{eqw}), for
$w$ leads to a dichotomy, depending on how the zero temperature is
reached.  For a straightforward $y\to\infty$ limit, there is an
unstable fp at $w_{\infty}^*={b}^{-1},$ separating the two stable
fp's at $0$ and $\infty$.  As $b>1$, $w_{\infty}^*<1$; it is in the
repulsive region.  It might appear to represent a dissociation of a three
chain bound state into a single strand and a bound pair at zero
temperature because of large three chain repulsion. It is to be noted
that $w=0$ is like hard core repulsion, preventing overlap of three
strands.

Curiously, the recursion relations also allow a different set of fixed
points for the relative weight
$$X=wy^2.$$  
In the $y\to\infty$ limit, the RG relation from Eq.~(\ref{eqw})
becomes
\begin{equation}
X'=\frac{3(b-1)}{b}+\frac{1}{b}X^2, \quad (y\to\infty),\label{eq:8}
\end{equation}
which has two fixed points
\begin{equation}
   \label{eq:3}
    X_{\pm}=\frac{b}{2}\pm\frac{1}{2}\sqrt{b^2-12(b-1)}.
\end{equation}
These fixed points are real for 
$b\ge b_X=2(3+\sqrt{6})=10.8989794..$, {\rm or} $b<2(3-\sqrt{6})=1.1...
.$ 
Of these, the latter one is not meaningful and therefore not
considered in this work.


A given system follows a path $w=y^{\eta/\epsilon}$ as $T$ is varied,
with $w\to 0$ or $\infty$ as $T\to 0$, depending on the sign of
$\eta$.  Therefore, the fixed point $w_{\infty}^*=1/b$ does not play
any role.  We instead focus on the zero temperature fixed points for
$X$.

\subsubsection{Justification of $X$}\label{sec:justification-x}
To see why $X$ is important, when the three chain weight is
$wy^3$, let us calculate the energy of the states.  If $n_2, n_3$
represent respectively pure two chain, and three chain (mutually
exclusive) contacts per unit length, then $n_2+n_3\le 1$.  The total
energy per unit length of DNA is
\begin{equation}
\frac{E}{N}=-n_2\epsilon+n_3(\eta-3\epsilon).
\end{equation}
On minimization,  
\begin{equation}\label{points}
\frac{E}{N}\Big|_{\rm min}=\left\{
\begin{array}{llr}
\eta-3\epsilon& {\rm if}\ \eta<2\epsilon &{\rm for} \ n_2=0, n_3=1,\\
-\epsilon& {\rm if}\ \eta>2\epsilon &{\rm for} \ n_2=1, n_3=0.
\end{array}
\right.
\end{equation}  
The zero temperature transition occurs when $\eta=2\epsilon$ with the
energy parameters as the variables. The high $\eta (>0)$ phase
consists of bubbles made of single chain and bound duplex, with
nonzero entropy.  Naively, if $\Delta S$ is the low temperature
entropy difference per bond of these two phases, one may combine it
with Eq.~(\ref{points}) to determine the free energy difference,
$\Delta F=\Delta E-T \Delta S$.  The continuity of the free energies
at the transition, i.e., $\Delta F=0$ then gives the transition
temperature as $wy^2\sim \exp(-|\Delta S|)$.

The above argument works for a first order transition as we see from
Eq.~(\ref{points}).  However local bubble formation in the triplex
bound state at low temperatures softens the system.  The ground state
energy is independent of dimensions.  As a result, there are two
mutually exclusive possibilities, viz., ({\it{i}}) no transition, or 
({\it{ii}}) a
continuous transition.  In absence of a good estimate of the bubble
entropy, this simple argument does not give a clue about the critical
dimension for the transition ($d>d_X\approx 4.46$, where $d_X= \ln(2
b_X)/\ln 2$, see below), but, in any case, it justifies the emergence
of the combination variable $X=wy^2$ in the low temperature region, as
noted in Sec. \ref{sec:zero-temp-fixd}.

$X$ is relevant for a phase transition from a three chain bound state
to a state of bubbles and single chain, while the three chain
Boltzmann weight $wy^3$ plays a role in melting phenomenon of the
Efimov state.
The bound-mixed transition is not a straight forward peeling
transition where the three chain bound state may split into a pair and
a single one.  If it were so, the transition would be equivalent to a
two chain melting case because at very low temperatures a pair would
be more-or-less bound acting like a single flexible polymer (in this
model).  The equivalent pairing interaction would be then $X=wy^2$,
yielding a peeling-off temperature $X=b-1$.  This is not the case, as
we see from Eq.~(\ref{eq:3}).  A comparison of Eq.~(\ref{eq:8}) with
Eq.~(\ref{eqy}) shows the difference.  The factor of 3 in the RG
equation for $X$ vis-a-vis Eq.~(\ref{eqy}) for $y$ shows the extra
entropic contribution coming from strand exchange.  This extra entropy
makes the bubble-bound phase a unique phase for three-stranded DNA.

\begin{figure}[htpb]
\includegraphics{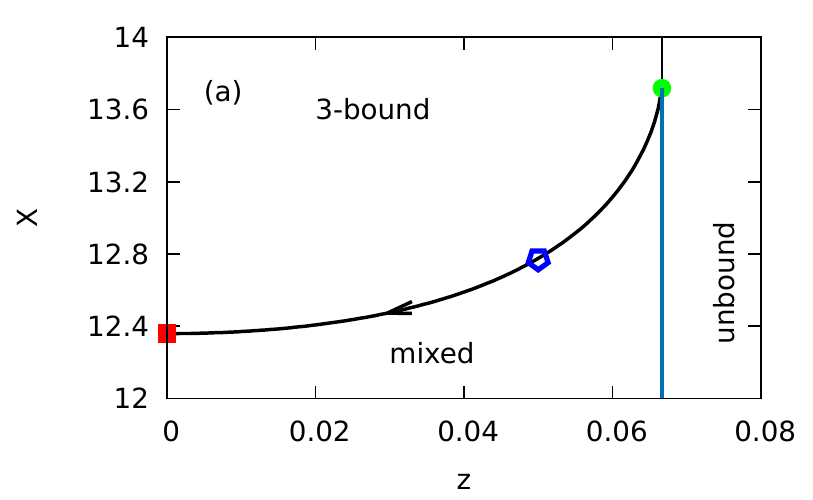}
\includegraphics{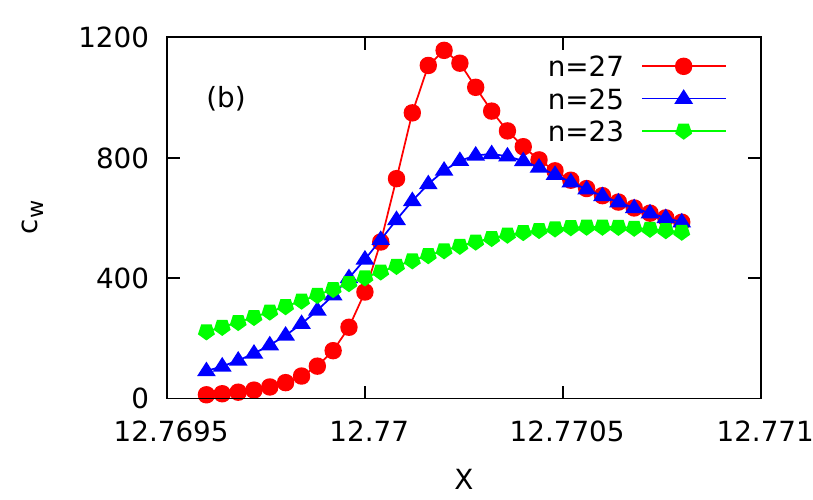}\\
\includegraphics{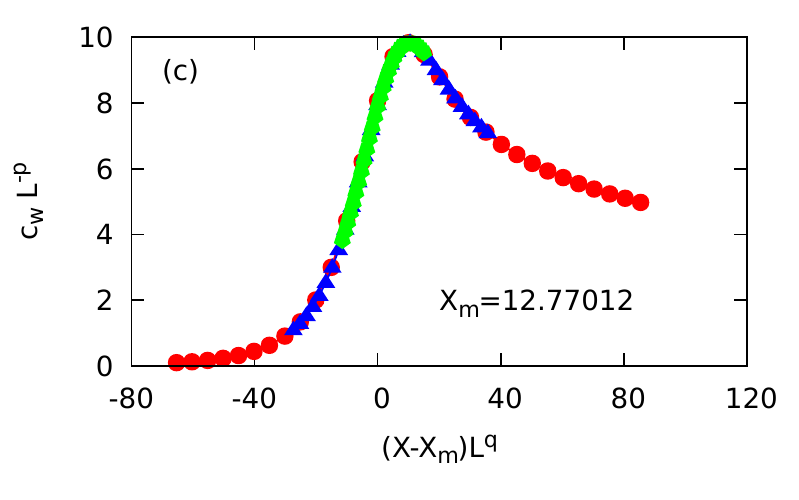}
\caption{(a) The separatrix in the $X$-$z$ plane, where $X=wy^2$ and
  $z=1/y$.  This is for $d=5$ which corresponds to $b=16$.  The two
  chain melting point is $y_c=15$.  The relevant region shown is for
  $0\leq z\leq z_c\equiv 1/y_c$.  The solid curve is the numerically
  determined separatrix connecting the unstable fixed points at
  $z=z_c$ (green filled circle) and at $z=0$ (red filled square).  The
  open pentagon (blue) represents the transition point obtained from
  the specific heat data in (b).  (b) Exact specific heat $c_w$ for
  fixed $y=20$, as a function of $X$ for three different lengths of
  chains, $L_n=2^n$ with $n=23,25,27$.  The size dependence indicates
  a diverging specific heat, and therefore, a continuous transition.
  (c) Finite size scaling: Same data as in (b) but plotted with scaled
  variables, with $p=\alpha/\nu, q=1/\nu$ and $X_m=12.77012$.  This
  value of $X_m$ at $z=0.05$ is shown by the open pentagon in (a).  }
\label{fig:separat}
\end{figure}  

\subsection{Phase diagrams for large $b$:  $y>y_c$}
\subsubsection{From RG: Separatrix}
The existence of real fixed points of $X$ for $y\to\infty$, allows us
to draw the phase diagrams for $b>b_X$.  This range of $b$ corresponds
to $d>4.446...$. Based on Eqs.~(\ref{pointt}) and (\ref{eq:3}), we have
three different situations.  ({\it{i}}) $b<b_c=8.56...$, ({\it{ii}}) 
$b_c<b<b_X$, and ({\it{iii}}) $b>b_X$.  In case ({\it{i}}), there are no fixed
points for $X$ for $y\geq y_c$.  The separatrix for the flow pattern
in the $y$-$w$ plane does not correspond to any transition since the
three chain bound state is the only thermodynamic phase.  Even though
there is a pair of fixed point for $y=y_c$ in case ({\it{ii}}), still there
is only the three chain bound phase.  Note that the fixed point for
$w$ at $w_{\infty}^*=1/b$ goes to infinity when expressed in terms of $X$.

The situation is different for $b>b_X$. The RG equations written in
terms of $X$ and $z=1/y$ are
\begin{subequations}
\begin{eqnarray}\label{eq4z}
   z'&=&\frac{bz^2}{(b-1)z^2+1}\approx b z^2 +O(z^4),\\
  X'&=& \frac{X^2 +3(b-1) + (b-1)(b-2) z^2}{b [(b-1) z^2+1]}, \label{eq:4} 
\end{eqnarray}
\end{subequations} 
which can be used to study the region $z<z_c=y_c^{-1}=(b-1)^{-1}$.  In
this low temperature regime, the unstable fixed points at $z=z_c$ and
$z=0$ are connected by a separatrix as shown in
Fig.~\ref{fig:separat}a for $b=16$.  In this case of $b=16$, the fixed
points are
\begin{subequations}
\begin{eqnarray}
  \label{eq:2}
X_{\pm}^{(c)}&=&13.7184, 3.34831,\quad {\rm at}\;\; y=y_c,\\
X_{\pm}&=& 12.3589, 3.6411, \quad {\rm at}\;\; y\to\infty. 
\end{eqnarray}
\end{subequations}
The global stable point $X_{-}$ represents the mixed phase.  The local
slopes of the separatrix at the fixed points are consistent with the
eigen-directions of the linearized versions of Eqs.~(\ref{eq4z}) and
(\ref{eq:4}) around the fixed points.  E.g.,
the horizontal tangent at $X_+$ follows from the absence of any linear
term in Eq.~(\ref{eq4z}) for small $z$. The flow along the separatrix
is towards the zero temperature (or infinite $y$) fixed point.  That
the separatrix represents the phase boundary is established below by
an exact computation for $b=16$ which corresponds to $d=5$.

It is interesting to note that under strong repulsion, the critical
state at $y=y_c$ is represented by a stable fixed point.

The behavior of specific heat as measured by the derivative of free
energy with respect $w$, is controlled by the fixed point $X_+$ at
$z=0$.  The relevant exponents are\cite{zeros}
\begin{subequations}
\begin{eqnarray}
  \nu&=&\frac{\ln 2}{\ln \left .\frac{dX'}{dX}\right|_{X\to X_{+}}}=\frac{\ln 2}{\ln
    \left[1+\sqrt{1-12\frac{(b-1)}{b^{2}}}\right]},\label{eq:9a}\\
 \alpha&=&2-\nu,  \label{eq:9}
\end{eqnarray}
\end{subequations}
where $\nu$ describes the divergence of an appropriate length scale,
while $\alpha$ describes the specific heat as $c_w\sim
|X-X_m(z)|^{-\alpha}$, $X_m(z)$ being the transition point.  The
specific heat here is defined as
\begin{equation}
  \label{eq:10}
c_w=-\frac{1}{L_n} \;w\frac{\partial\ }{\partial w} w \frac{\partial \ln Q}{\partial w},
\end{equation}
keeping $z$ constant. By construction, $c_w$ is related to the
fluctuations in the number of three chain contacts.  Eq.~(\ref{eq:9})
shows that there is a continuous transition for all $d>4.46$ but no
divergence in specific heat for $b<13.4$, i.e. $d<4.74$.

\subsubsection{Exact computation of $c_w$}
The specific heat or the fluctuations in the number of three contacts
can be computed exactly for finite lengths in a recursive scheme.  The
recursion relations for the partition functions are given by
\begin{subequations}
  \begin{eqnarray}
    C_{n+1}&=&b^2 C_n,    \label{eq:5}\\
    Z_{n+1}(y)&=& b Z_n(y)^2 + b(b-1)C_n^4,\label{eq:6}\\
    Q_{n+1}&=& b Q_n^2 +b(b-1)Z_n^2 C_n^2\nonumber\\
           &&\quad + b(b-1)(b-2) C_n^6,\label{eq:7}
  \end{eqnarray}
\end{subequations}
where $C_n, Z_n, Q_n$ are the $n$th generation partition functions for
single, double and triple strand cases.  One may also write down the
recursion relations for the derivatives.  To be noted here that the
three chain interaction $w$ does not affect $Z$, and therefore the
recursion relation for $c_w$ is simpler than other derivatives.

By iterating the recursion relations, $c_w$ has been computed exactly
upto $n=27$.  See Fig.~\ref{fig:separat}b where $c_w$ is plotted as a
function of $X$ for $y=20$. The strong growth with size is an
indication of a diverging specific heat. A finite size scaling form
suggests that all these data points can be collapsed on to a single
curve if plotted as $c_w L_n^{-p} $ {\it vs} $(X-X_m) L_n^{q}$, with
$p=\alpha/\nu, q=1/\nu$, where $\nu$ and $\alpha$ are given by
Eqs.~(\ref{eq:9a}) and (\ref{eq:9}).  Here, $L_n=2^n$.  In this way
$X_m(z)$ can be estimated by using the data collapse measure of
Ref.~\cite{datcol}.  The data collapse is shown in
Fig.~\ref{fig:separat}c.  The transition point sits nicely on the
separatrix in Fig.~\ref{fig:separat}a.

\mixedsch

\subsection{Summary for $d>2$}
The end result is that for large $d>d_X$, there is now a new phase,
the bubble-bound phase or the mixed phase, at low temperatures in
presence of three chain repulsion.  A possible form of this
three-stranded DNA with pairwise bound but without three chain
contacts is shown by the schematic diagram in Fig.~\ref{fig:mix-sch}.
In the mixed phase two are bound with one free over a certain length
scale of the chain, but the strand exchange mechanism leaves no one
free completely.  This phase undergoes a continuous transition to a
completely bound state, where the two chain attraction overcomes the
weak local three chain repulsion.


For easy reference we show all possible states schematically in
Fig.~{\ref{fig:dis}} for various dimensions.  Three typical cases
shown are (a) $b=8$ (i.e. $d=4$), (b) $b=9$ ($d=4.1699$), and (c)
$b=16$ ($d=5$).  Fig.~{\ref{fig:dis}}a is similar to the $b=4\, (d=3)$
case of Ref. \cite{zeros}.  The phase diagram displays the Efimov
region and the mixed phase in addition to the conventional bound and
the unbound states.  The Efimov region is defined as the region in the
3-chain bound phase, where no two are supposed to be bound or the
three not bound by $w$ alone.  In other words, this is a state where
we would not see a bound state if either a chain is removed, or there
is no two body attraction (i.e., $\epsilon=0$).  The traditional
Efimov case corresponds to $w=1, y<y_c$.  However, there is no well
defined thermodynamic boundary for the Efimov region.  So in the
$w$-$y$ plane, the Efimov region lies inside the domain of the triplex
phase with $w<w_c$ (pure three chain melting at $y=1$) and $y<y_c$
(duplex melting).  Such regions enclosing the Efimov states are marked
as ``Efimov'' in Fig.~{\ref{fig:dis}}.

\discuss

The mixed phase appears under the thin curve in Fig.~\ref{fig:dis}c
for $b>b_X$.  The transition line between the unbound and the Efimov
state is first order.  The vertical melting line for the mixed to
unbound phase and the mixed to bound phase transitions are both
continuous.  The transition from the mixed bubble-bound to the bound
state is associated with diverging specific heat for high enough $d$.

We see that when there is an unstable fp at $y=y_c$, the separatrix
connecting $(y=y_c,X=X_+^{(c)})$ and $(y=1,X=w_c)$ defines the Efimov
line (represented by the thick line).  Thus, $d=4$ is very special
(Fig.~{\ref{fig:dis}}a) among the three cases shown, where an
Efimov-DNA occurs for any $w$ at $y=y_c$.  In the intermediate range
of dimensions, say $4.0976...<d<4.446...$, there is no mixed phase.
The phase diagram (Fig.~{\ref{fig:dis}}b) resembles
Fig.~{\ref{fig:dis}}c but without the mixed phase and the curved
transition line.  Curiouser and curiouser here is the difference in
the melting behavior of the three chain bound state for $X>X_+^{(c)}$
and $X<X_+^{(c)}$.  For $X>X_+^{(c)}$, the three chain bound state
melts via the Efimov line, a first order melting \cite{zeros}, but for
$X<X_+^{(c)}$, there is a continuous melting identical to the two
chain melting problem.

Let us list the lower critical dimensions ($d_{\rm lc}$) for the
various phases (or transitions) we see in DNA:
\begin{enumerate}
\item dsDNA melting: $d_{\rm lc}=2$.
\item Pure tsDNA melting: $d_{\rm lc}=1$.
\item Melting of Efimov-DNA: $d_{\rm lc}=2$.
\item Existence of bubble-bound mixed phase: $d_{\rm lc}=d_X\approx
  4.446$ (see below Eq.~(\ref{eq:3})).
\item Critical melting of tsDNA (under repulsion): $d_{\rm
    lc}=d_c\approx 4.0976$ (see below Eq.~(\ref{eq:1})).
\end{enumerate}

\section{Model: $d<2$}\label{model2}
Further to verifying the validity of such a finding, the mixed phase
in particular, we consider a similar model on a lower dimensional
lattice, namely the Sierpinski gasket. 
See Fig.~\ref{fig:latsg} for generations $n=0,\ 1$.  It is for these
low dimensional cases where the mixed phase was first identified.  The
construction starts from a unit triangle, which is repeated
iteratively for each triangle in a self-similar way to form a bigger
lattice.

\sglattice 
\subsection{Chains on a Sierpinski Gasket}
The polymers are restricted to the non-horizontal bonds of a
Sierpinski Gasket to keep DNA length same for all the strands.  The
DNA problem can be classified into two different cases according to
the constraints on walks.  In one case polymers cannot cross each
other and in the other case polymers can cross each other at any
length of the polymers.  The triple-chain melting was discussed in
ref\cite{mixed}, where weights $\sigma_{ij}$ and $\sigma_{ijk}$ were
assigned at the vertex for bubble opening or closure, and $y$ for
sharing same bond by two strands.  Here we define a more general model
by assigning an extra weight $w$ for three chains sharing a bond. As
in Sec. II, if three strands share a bond, the weight is $y^3 w$.
This model contains some of the models of Ref.~\cite{mixed} and it
reproduces essentially all of the results discussed there, in
appropriate limits.

\lattschsg

\subsection{Partition functions}
To describe exactly all possible configurations of the three chain
system, we need to introduce the following partition functions,
$a_{10}$, $a_{11}$, $a_{20}$, $a_{21}$, $a_{22}$, $a_{30}$, $a_{31}$,
$a_{32}$, $a_{33}$.  Different possible polymer walks are shown in
Fig.~\ref{fig:lat-schsg}.  The generating functions in terms of sum
over all the configurations for the non-crossing case can be written
as
\begin{subequations}\label{eqrec}
\begin{align}
  A_{10}&= a_{10}^2,\\
  A_{20}&= a_{20}^2,\\
  A_{30}&= a_{30}^2,\\
  A_{11}&= a_{10}^2\ a_{11} +  a_{11}^2,\\
  A_{21}&= (a_{10}\ a_{11} +  a_{01}\ a_{20})\ a_{21},\\ 
  A_{31}&= (a_{11}\ a_{20} +  a_{01}\ a_{30})\ a_{31},\\
  A_{22}&= a_{11}\ a_{21}^2 + a_{20}^2\ a_{22} +  a_{22}^2,\\
  A_{32}&= a_{21}^2\ a_{31} + (a_{10}\ a_{22} +  a_{02}\ a_{30})\ a_{32},\\
  A_{33}&= a_{22}\ a_{31}^2 + a_{11}\ a_{32}^2 + a_{30}^2\ a_{33} +  a_{33}^2,  
\end{align}
\end{subequations}
with the initial conditions
\begin{eqnarray} \label{inicond}
&a_{10}=1, a_{11}=1, a_{20}=y, a_{21}=y, a_{22}=y^2, \nonumber \\
&a_{30}=y^3w, a_{31}=y^3w, a_{32}=y^4w, a_{33}=y^6w^2.       
\end{eqnarray} 
Here we have set the bubble initiation factors
$\sigma_{ij}=\sigma_{ijk}=1$.  This is a generalization of the TS1
case of Ref.~\cite{mixed}.

The generating functions for the crossing case can be written as (rest
are same as Eq.~\ref{eqrec})
\begin{subequations}\label{eqrecc}
\begin{align}
  A_{22}&= a_{11}\ a_{21}^2 + 2 a_{20}^2\ a_{22} +  a_{22}^2,\\
  A_{32}&= 2 a_{21}^2\ a_{31} + (a_{10}\ a_{22} +  a_{02}\ a_{30})\ a_{32},\\
  A_{33}&= 3 a_{22}\ a_{31}^2 + 3 a_{11}\ a_{32}^2 + a_{30}^2\ a_{33} +  a_{33}^2,  
\end{align}
\end{subequations}
with the initial conditions shown in Eq.~\ref{inicond}.  

The total partition functions for the two chain and the three chain systems
are given by 
\begin{equation}\label{eqZ}
Z_{2}= A^2_{11}+A_{22},
\end{equation}
\begin{equation}\label{eqXi}
  Z_{3}= A^3_{11}+A_{11}A_{22}+A_{33},
\end{equation}
with their logarithms giving the free energies.  The two terms in
Eq.~\ref{eqZ} are for the unbound and for bound chain configurations.
The first term of the two chain equation dominating over the second
one results in a phase transition.  A similar procedure of comparison
has been adopted for the three chain system.  The unbound, the bound,
and the mixed phases appear over a wide range of temperatures.  The
mixed phase is represented by the middle term of Eq.~\ref{eqXi}.  To
obtain the phase diagram in $w$-$y$ plane we look for the convergence
or divergence of the following ratios
\begin{equation}\label{eqrat}
r_1=\frac{A_{22}}{A^2_{11}}, \ r_2=\frac{A_{33}}{A^3_{11}}, \;\; {\rm and}\;\;\ r_3=\frac{A_{33}}{A_{11}A_{22}}.
\end{equation}
The phase boundaries separating the phases can be obtained from
condition $r_i=1$, for any $i=1,2,3$.

\mixedsgnc

\subsection{Phase diagrams}
\subsubsection{Non-crossing case}
The phase diagram for the non-crossing case is shown in
Fig.~\ref{fig:mix-sgnc}. All the three phases, three-bound, 
three-unbound and mixed, appear here.  The vertical line $y=y_c$ corresponds
to the two chain melting and it is also the melting line for the mixed
phase.  On the bound side, $y>y_c$, there is a triplex-mixed phase
transition line.  The topology of the phase diagram is similar to
Fig.~\ref{fig:dis}c, but here all are first order lines.  The three
transition lines meet at the triple point at $y_c$.  The dashed line,
$w=1/y$ verifies the results of TS1 model of Ref.~\cite{mixed}.  TS1
is defined as the model of noncrossing walks with $y_{12}=y_{23}=
y_{31}=y$, $\sigma_{ij}=\sigma$, and $w=1/y$.  A sequence of
transitions
$${\rm triplex \leftrightarrow mixed \leftrightarrow denatured}$$
occurs in this case.

\mixedsgc

\subsubsection{Crossing case}
For $y>y_c$ when chains are supposed to be in pairs, the three chain
repulsion plays an essential role to produce the mixed phase by
stopping three monomers contact at a time.  For the non-crossing case
each of the three phases (bound, unbound, and mixed) occurs for a wide
range of temperatures whereas for the crossing case there is no
two chain melting at any finite temperature.  See
Fig.~\ref{fig:mix-sgc}.  This is because the bubble entropy in low
dimensions is not enough to induce a melting. As a result the DNA
strands remain bound at all temperatures for any arbitrarily weak
short range pair attraction.  Nevertheless the situation can be
changed if the three chain repulsive interaction is incorporated.  A
transition from the bound to the mixed state is possible by tuning the
three chain repulsive force among chains.

Furthermore transitions can be induced in the crossing case if we
introduce the bubble initiation or closure factor $\sigma$ in
Eq.~\ref{inicond}, {\it e.g.}, by considering the initial conditions
$a_{21}=y \sigma$, $a_{31}=y^3w \sigma^2$, $a_{32}=y^4w \sigma^2$, as
in the TS2 model of Ref.~\cite{mixed}.

\mixedsgp 

With $\sigma=0$ when the bubble formation is suppressed, the chains
will be either open or bound, {\it i. e.} once the chains are unbound
reunion is not possible along the length of chains.  This is the
Y-fork model.  In the Y-fork model the crossing and the non-crossing
case do not have any difference as the strand exchange is not allowed
for $\sigma=0$.  We refer to Ref.~\cite{mura} for discussions on the
Euclidean lattice results.

\sigyw

By equating the partition function of the bound state to the partition
function of the unbound state, the duplex melting turns out to be
\begin{equation}
  A_{11}y^N=A_{11}^2 \implies y_0=A_{11}^{1/N}, 
\end{equation}
where  $y_0=y_c(\sigma=0)$.  It is known that $y_0=1.26408....$.  
We may use the same logic for the triplex melting  
\begin{equation}\label{wt}
  w_t=\frac{A_{11}^{2/N}}{y^3}=\frac{y_0^2}{y^3}
\end{equation}
This relation fits the numerically obtained points in
Fig.~\ref{fig:mix-sg1}, suggesting that the phase is a triple chain
bound state and not necessarily the Efimov-DNA.  This should be the
case because with $\sigma=0$ there are no bubbles.  This point is
further elaborated below.
 
Using similar logic as above, the triplex to mixed transition occurs
at
\begin{equation}\label{ym}
  w_m=\frac{A_{11}^{1/N}}{y^2}=\frac{y_0}{y^2}.   
\end{equation}
where $N=2^{n+1}$ is the length of a polymer.  The arguments used here
are very similar to those in Sec. \ref{sec:justification-x}, and show
the roles played by $wy^3$ and $wy^2$ in different transitions.

All the above three transitions occur at the same temperature $y=y_0$
if we choose $w=1/y$ in a Y-fork model.  It is also apparent from
Fig.~\ref{fig:mix-sg1} that the three chain phase boundary and the
mixed phase boundary meet at $w_0=1/y_0$ for $\sigma=0$.

Bubbles are essential for the Efimov-DNA and the bubble-bound mixed
phase.  Therefore, Efimov-DNA may occur for $\sigma\neq 0$ but
definitely not at $\sigma=0$.  The triplex melting line $w_t$ given by
Eq.~\ref{wt} is from the tightly bound to the unbound phase, without
any bubble.  Incidentally, the Efimov state is a continuation of the
three chain bound phase, and not a distinct thermodynamic phase; there
is no phase boundary to protect it, except for melting.  We may then
identify the Efimov region for $0<\sigma<1$ as the region $w<w_t$ upto
the corresponding melting line and $y<y_c(\sigma)$ (see
Fig.~\ref{fig:3dplot}).  This is the region where the bubbles
contribute most.  A similar situation arises in the context of the
bubble-bound mixed phase, whose existence is also at stake at
$\sigma=0$.  In the $\sigma=0$ limit, a transition takes place at
$w_m$ (Eq.~\ref{ym}) to a phase where any two will be bound throughout
and one free (see Fig.~\ref{fig:mix-sch}).  Absence of bubbles
strictly imply no strand exchange.  As a result, the transition is
actually like peeling off one chain from the three.  We may still call
this a mixed phase by continuation because, by symmetry, any one can
be unbound.

The projection of the $w$-$y$-$\sigma$ phase diagram for the crossing
case is depicted in Fig.~\ref{fig:3dplot} for different $\sigma$ ($=1,
0.8, 0.6, 0$) in the $w$-$y$ plane.  The melting lines for the
three chain and the mixed phase meet at $w_t(\sigma)=w_m(y_c)$ for any
given value of $\sigma$.  Model TS2 of Ref.~\cite{mixed} can be
recovered exactly from this model.  The curve $w=1/y$ intersects the
mixed phase boundaries for different values of $(\sigma,
y_c(\sigma))$.  These data certainly reproduce the $\sigma$-$y$ phase
diagram of the TS2 model in Ref.~\cite{mixed}.

\section{conclusion}
\label{modeldis}

The role of dimensionality of the underlying lattices on the emergence
of the mixed or the bubble-bound phase of a triple-stranded DNA has
been the focus of this paper.  The mixed phase was first discussed in
the context of a class of DNA models on low dimensional fractal
lattices, where a melting transition is induced by a bubble initiation
factor that suppresses bubble entropy.  We showed that the phase
remains stable even with three body repulsion.  With native DNA pair
interaction and three chain repulsion, we further showed that such a
phase is thermodynamically stable on the bound side of a duplex if
dimensionality is large ($d>4.5$). We established a diverging specific
heat for $d=5$ from both renormalization group and finite size scaling
analysis of exact computations.

When there exists a bubble-bound phase, the topology of the phase
diagram remains the same for both higher and low dimensional models.
However the nature of transitions are different.  In general, the
transition from the mixed bubble-bound phase to the three chain bound
state is continuous for $d>4.5$, but for $d<2$, all transitions are
first order.

\end{document}